\documentclass[conference]{IEEEtran}

\usepackage{courier}        
\usepackage{type1cm}        

\usepackage{makeidx}         
\usepackage{graphicx}
\usepackage{wrapfig}
\graphicspath{ {images/} } 

 \usepackage[justification=raggedright]{subcaption}
\usepackage{courier} 

\usepackage{multicol}        
\usepackage{multirow}
\usepackage{rotating} 
\usepackage[bottom]{footmisc}
\usepackage[usenames,dvipsnames,svgnames,table]{xcolor}
\usepackage{amsmath}
\usepackage{float}
\usepackage{todonotes}
\usepackage{import}

\usepackage{booktabs} 
\usepackage[flushleft]{threeparttable}
\usepackage[draft=false,colorlinks=true,linkcolor=blue]{hyperref} 
\usepackage{enumitem} 
\usepackage[numbers,sort&compress,square]{natbib} 

\usepackage{fancyhdr}
\usepackage{kantlipsum}
\fancypagestyle{plain}{
\fancyhf{}
\fancyhead[C]{Conference on \LaTeX} 

}
\usepackage{eso-pic} 

\IEEEoverridecommandlockouts 

\begin{document}

\AddToShipoutPictureBG*{
\AtPageUpperLeft{
\setlength\unitlength{1in}
\hspace*{\dimexpr0.5\paperwidth\relax}
\makebox(0,-0.75)[c]{\textbf{2018 IEEE/ACM International Conference on Advances in Social Networks Analysis and Mining (ASONAM)}}}}

\title{\LARGE{EGBTER: Capturing degree distribution, clustering coefficients, and community structure in a single random graph model}\thanks{\scriptsize{This manuscript has been authored by UT-Battelle, LLC under Contract No. DE-AC05-00OR22725 with the U.S. Department of Energy. The United States Government retains and the publisher, by accepting the article for publication, acknowledges that the United States Government retains a non-exclusive, paid-up, irrevocable, world-wide license to publish or reproduce the published form of this manuscript, or allow others to do so, for United States Government purposes. The Department of Energy will provide public access to these results of federally sponsored research in accordance with the DOE Public Access Plan (\url{http://energy.gov/downloads/doe-public-access-plan}).}}
}

\author{
\IEEEauthorblockN{Omar El-daghar}
\IEEEauthorblockA{Department of Mathematics\\
Purdue University\\
West Lafayette, IN 47907\\
oeldaghar2013@gmail.com}
\and
\IEEEauthorblockN{Erik Lundberg}
\IEEEauthorblockA{
Department of Mathematics\\ 
Florida Atlantic University\\
Boca Raton, FL 33431\\
elundber@fau.edu}
\and
\IEEEauthorblockN{Robert Bridges}
\IEEEauthorblockA{
Computational Sciences \& Engineering Division\\
Oak Ridge National Laboratory\\
Oak Ridge, TN 37831\\
bridgesra@ornl.gov}
}

\maketitle

\IEEEoverridecommandlockouts
\IEEEpubid{\parbox{\columnwidth}{\vspace{8pt}
\makebox[\columnwidth][t]{\textbf{IEEE/ACM ASONAM 2018, August 28-31, 2018, Barcelona, Spain}}
\makebox[\columnwidth][t]{\textbf{978-1-5386-6051-5/18/\$31.00~\copyright\space2018 IEEE}} \hfill} \hspace{\columnsep}\makebox[\columnwidth]{}}
\IEEEpubidadjcol


\begin{abstract}

Random graph models are important constructs for data analytic applications as well as pure mathematical developments, as they provide capabilities for network synthesis and principled analysis. 
Several models have been developed with the aim of faithfully preserving important graph metrics and substructures. 
With the goal of capturing degree distribution, clustering coefficient, and communities in a single random graph model, we propose a new model to address shortcomings in a progression of network modeling capabilities.  
The Block Two-Level Erd{\H{o}}s-R{\'e}nyi (BTER) model of Seshadhri et al., designed to allow prescription of expected degree and clustering coefficient distributions, neglects community modeling, while the Generalized BTER (GBTER) model of Bridges et al., designed to add community modeling capabilities to BTER, struggles to faithfully represent all three characteristics simultaneously. 
In this work, we fit BTER and two GBTER configurations to several real-world networks and compare the results with that of our new model, the Extended GBTER (EGBTER) model. 
Our results support that EBGTER adds a community-modeling flexibility to BTER, while retaining a satisfactory level of accuracy in terms of degree and clustering coefficient. 
Our insights and empirical testing of previous models as well as the new model are novel contributions to the literature.

\end{abstract} 
\section{Introduction}
\label{sec:2}
Graphs provide natural representations of objects and their relationships, and are now a prevalent tool in the applied sciences, including
biology, physics, and computer science~\cite{aittokallio2006graph, goldenberg2010survey, pirzada2007applications, prvzulj2009computational, karlan2009trust, liu2015graph, dawood2014graph, dunn2012identifying, musolesi2007designing, bansal2007infer, castro2002scribe, amiri2017snapnets}. 
Random graph models, which prescribe probability distributions over a set of graphs,
have become a mainstay for both applied and pure graph theory~\cite{erdos1960evolution, chung2002average, watts1998collective, holland1983stochastic, seshadhri2012community, mahadevan2006systematic, bridges2015multi, bridges2016multi}.
They give probabilistic machinery admitting principled network simulation capabilities that facilitate statistical analysis in the face of data that is scarce, time-consuming, or impractical to apprehend.
Complex networks, especially those arising from natural representations of real world data, exhibit structural idiosyncrasies,  which are measured and deduced using many heterogeneous graph metrics, e.g., distributions of degrees, clustering coefficients, and motifs as well as measures of size, centrality, and modularity, to name a few. 
Consequently, a major focus in random graph models has been on identifying those graph metrics that are important in real-world graph representations and on developing models that preserve these metrics \cite{airoldi2008mixed, barabasi1999emergence, chung2002average, chung2002connected, eberle2007anomaly, kolda2014scalable, mahadevan2006systematic, miller2012goodness, moreno2013network, seshadhri2012community, seshadhri2013triadic,sheng2011undetermination,kajdanowicz2016using}. 

This paper identifies and addresses gaps in a progression of graph models presented recently in the research literature. 
Building on the famous Erd\"{o}s-R\'{e}nyi (ER) and Chung-Lu (CL) models, the Block Two-Level Erd\"{o}s-R\'{e}nyi (BTER) introduced by Seshadhri et al. (2012)~\cite{seshadhri2012community} was designed to allow specification of degree distribution and clustering coefficient per degree. 
For understanding social networks, clustering coefficient is an important metric as it is driven by the nature of real-world relationships. 
Initial tests show accurate modeling of clustering coefficients per degree as well as the degree distribution by BTER. 
Further work of Kolda et al. \cite{kolda2014scalable} discuss scalable implementations of BTER.

Similarly, understanding community structure (number, size, within- and between-community densities) and membership  (who i.e., which nodes, interacts with whom) is important for network analysis, as it shows important relationships and their changes. 
As BTER does not allow specification of communities, Bridges et al. (2015) introduced a Generalized BTER model, (GBTER)~\cite{bridges2015multi,bridges2016multi} adding user flexibility to prescribe community membership and community edge densities, 
while retaining the flexibility to also prescribe (most) degree distributions in expectation. 
As initial efforts with GBTER focused on aiding anomaly detection algorithms (specifically, GBTER's probabilistic formulation allowed p-value computations, which were used for identifying anomalous changes in node degree and community membership in time-varying graphs), GBTER's efficacy in modeling communities while preserving degree or clustering coefficient properties is unknown. 

In this work we investigate how well GBTER configurations model many real world networks. 
We identify shortcomings of this model and provide quantitative results and explanations for why the model fails to preserve desired characteristics. 
In light of these findings, we introduce an improved model, the Extended GBTER  (EGBTER), and exhibit it's modeling capabilities on the same set of networks. 
An advantage of the GBTER and EGBTER contributions (as well as Stochastic Block Model \cite{holland1983stochastic}) is that one can model not just community structure (in terms of number of communities, size and density of each, and interactions between communities) but also specify membership. 

Our contributions provide qualitative comparisons of the BTER model and two configurations of the GBTER model, as well as quantifiable head-to-head results by using three graph metrics that reflect edge, clustering, and community structure statistics on several real-world networks. 
Our testing gives insight to the modeling capabilities of these relatively new models, and, broadly speaking, shows that BTER often succeeds in faithfully reproducing clustering metrics at the expense of community structure, while GBTER performs conversely. 

Informed by our analysis of BTER and GBTER, we introduce the EGBTER model, a natural combination of BTER and GBTER, addressing the limitations of both. 
More specifically, EGBTER is crafted to circumvent degree distribution problems within user-defined communities that are caused by the simplicity of the ER process used in the GBTER model, but retains the ability to specify community membership and density. 
We compare EGBTER both qualitatively and quantitatively to the previous two models. Our results suggest that EGBTER provides more balanced modeling of degree distribution, clustering coefficient, and community structure.

We note that many other random graph models seek to preserve similar characteristics.  
E.g., the Stochastic Block Model \cite{holland1983stochastic} seeks to faithfully preserve density between specified nodes.
The $dK$-Orbis models \cite{mahadevan2006systematic} prescribe distribution of edges between $d$-tuples of node degrees and have exhibited preservation of community structure and degree distribution for large enough $d$ (but do not allow specification of which nodes constitute each community). 
Comparison of EGBTER against other models besides BTER and GBTER is necessary for greater understanding, but is outside of scope for this work. 

\begin{table}
  \centering
  \vspace{-.1cm}
  \begin{threeparttable}
    \caption{Notation and metrics used throughout.}
        \begin{tabular}{p{.43\linewidth}p{.46\linewidth}}
            Notation & Description \\
            \toprule
        

        $ V$ & set of nodes \\
        $ E$ & set of edges \\
        $v_i$ & $i^{th}$ vertex\\
        $G[X]$ & vertex induced subgraph of $G$ on $X\subset V$\\
        $d_i$ or $deg_G(v_i)$ & degree of $v_i$ in graph $G$\\
        $\lbrace d \rbrace$ & degree distribution\\
        $n_d$ & number of nodes of degree $d$\\
        $RMSE \lbrace d\rbrace$ & root mean squared error of degree distribution\\
        $cc(v_i)=\frac{2L_i}{d_i(d_i-1)}$ & $v_i$ local clustering coefficient (CC)\tnote{1}\\
        $cc_d=\frac{1}{n_d}\sum_{d_i=d} cc(v_i)$ & average local CC for nodes of degree $d$ \\
        $\lbrace cc_d \rbrace=\lbrace (d,cc_d)\rbrace$ & CC per degree (CCPD) distribution\tnote{2} \\
        $\lbrace cc_{d,C_k} \rbrace$ & $cc_d$ of $G[C_k]$\\
         $RMSE \lbrace cc_{d}\rbrace$ &  root mean squared error of CCPD distribution\\
         $\lbrace cc_{d,C_k} \rbrace$ & $G[C_k]$ CCPD distribution\tnote{3}\\
         $CL(w_i,w_j)=\sum_k\frac{w_iw_j}{\sum w_k}$& Chung-Lu probability of adding edge $v_iv_j$ \\
         $C_k$& $k^{th}$ community of $G$\\
         

        $Q=\sum_i(e_{ii}-a_i^2)$ & modularity\tnote{4} \\
        $A_L$ & $L^{th}$ grouping formed by BTER model \\
        $\epsilon_i=\max\lbrace 0, d_i-(\vert C_k\vert-1) p_k\rbrace$ & $v_i\in{C_k}$ expected excess degree after ER process\tnote{5}\\
        $E_i=D_i-d_i$ & $v_i$ expected excess degree after within-$C_k$ CL process\\
        
            \bottomrule
        \end{tabular}%
        
    \begin{tablenotes}\footnotesize
        \item[1] $L_i$ is the number of links among neighbors of $v_i$. Quantity $cc(v_i)$ is the average probability that two neighbors of $v_i$ are also neighbors of one another.
        \item[2] CCPD distribution is the set of tuples $(d,cc_d)$. If there are no nodes of degree $d$ then $cc_d=0$.
        \item[3] This is the CCPD distribution of the induced subgraph $G[C_k]$.
        \item[4] Note that $e_{ij}$ is the fraction of total egdes that connect nodes in community $C_i$ to community $C_j$ and $a_i=\sum_j e_{ij}$.
        \item[5] With $d_i$ the expected degree of $v_i$, $\epsilon_i$ is the remaining expected degree of $v_i$ after generating edges from the internal ER process on $C_k$.
    \end{tablenotes}
    
    \label{tab:notation}%
    \end{threeparttable}
    \vspace{-.3cm}
\end{table}%

\section{Notation \& Previous Models}
\label{sec:3}

We use classical terms and notation except in
the case of \emph{clustering coefficient}, where we use \emph{average local clustering coefficient} in place of clustering coefficient of a node. We shall deal exclusively with simple graphs.  
See Table~\ref{tab:notation} for notation and definitions used throughout. 
We use \emph{density} to refer to the percentage of total possible edges present in a network. A triangle refers to a complete graph on three nodes. In the literature, notions of community and community structure vary. In this work, we use the algorithm of Louvain et al. \cite{blondel2008fast} to produce a partition of $V$. We refer to each set of nodes in this partition as a community and the partition itself as the community structure.

Next, we describe the previous models, BTER and GBTER. 
Both are built on combinations of two influential and historical models, ER and CL, which are described in the \hyperref[app:appendix]{Appendix}. 

\begin{figure}[b]
    \vspace{-.2cm}
    \includegraphics[width=\linewidth]{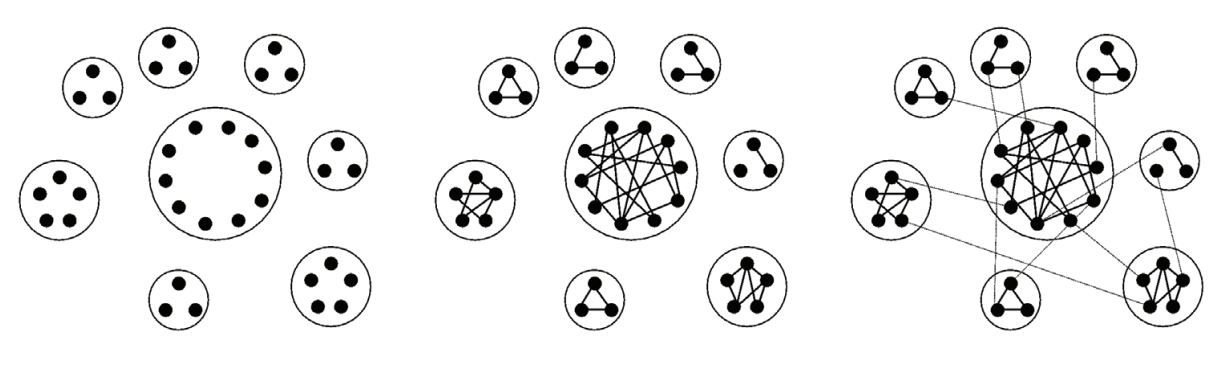}
    \caption{\footnotesize{BTER process depicted, (Left to right) nodes partitioned into  groupings $A_L$, within-group edges generated from ER process, edges from CL process generated. Figure from Seshadhri et al.~\cite{seshadhri2012community}.}}
    \label{fig:3}
    \vspace{-.2cm}
\end{figure}

\textbf{BTER Model:} The BTER model uses a two-step edge insertion process. 
In the first stage, nodes are partitioned based on the given expected degree sequence, and an ER model is sampled in the subgroups with parameter chosen to preserve the clustering coefficient per degree (CCPD). 
Next a CL model is applied to ensure the expected degree sequence is attained.\\

\noindent \textit{Inputs:}\\
(1) Expected degree distribution $\lbrace d \rbrace$,\\
(2) Expected CCPD  $\lbrace cc_d\rbrace$.\\
\textit{Generative process:}
\begin{enumerate}[leftmargin=.6cm]
    \item Node groupings are implicitly assigned by putting nodes of degree $d$ into groups of size $d+1$ and denoted by $A_L$. 
    \item Build an ER graph in each $A_L$ of size $d+1$ nodes with $p_{A_L}=cc_d^{1/3}$.
    \item Build a CL graph on the entire network with node weights $w_i : =\max \lbrace 0, d_i-cc_d^{1/3} (\vert C_k \vert-1) \rbrace$. 
    
\end{enumerate}
See Fig.~\ref{fig:3} for a visual depiction of the edge generation process. 
For a more detailed description, see the works of Kolda et al. and Seshadhri et al.~\cite{kolda2014scalable,seshadhri2012community}.

{\bf GBTER Model:} The Generalized BTER (GBTER) model was introduced by Bridges et al. \cite{bridges2015multi, bridges2016multi} and arose with a goal of detecting multi-level anomalies in a sequence of time-varying graphs. 
It generalizes the BTER model by giving the user flexibility to partition the nodes into groupings (to model communities if desired, and specifically which nodes participate in each community) and to specify the ER-density parameter of each community (e.g., to model within-community density or within-community clustering coefficient).  
Both are implicitly defined in the first two stages of the BTER generation process, but explicitly prescribed as inputs for GBTER. 
GBTER then uses a CL model to attain the expected degree sequence. 
The edge generation below combines both steps (ER then CL) into a single probability for each edge.
Setting the GBTER partition and ER density parameter to be that of BTER reduces the GBTER configuration to that of BTER.\\
\noindent\textit{Inputs:}
\begin{enumerate}[leftmargin=.6cm]
    \item Expected degree distribution, $\{d_i: v_i\in{G}\}$
    \item A community partition $P=\dot{\bigcup}_k C_k$ (community membership specified)
    \item Expected density parameter, $p_k$ for each community $C_k$ (modeling within-community density or within-community CCPD)
\end{enumerate}
\textit{Generative process:}
\begin{enumerate}[leftmargin=.6cm]
    \item Each edge is added with probability
    \begin{equation}
        P(v_iv_j\in{E})=\begin{cases}
        p_k+(1-p_k)CL(\epsilon_i,\epsilon_j), & \text{if } v_i,v_j\in{C_k}\\
        CL(\epsilon_i,\epsilon_j), & \text{otherwise,}
        \end{cases}
        \label{eqn:gbter}
\end{equation} where $\epsilon_i=\max{\lbrace 0, d_i-p_k(\vert C_k \vert -1)}\rbrace$.
\end{enumerate}

\begin{figure}
    \centering
    \vspace{-.4cm}
    \includegraphics[width=0.4\textwidth]{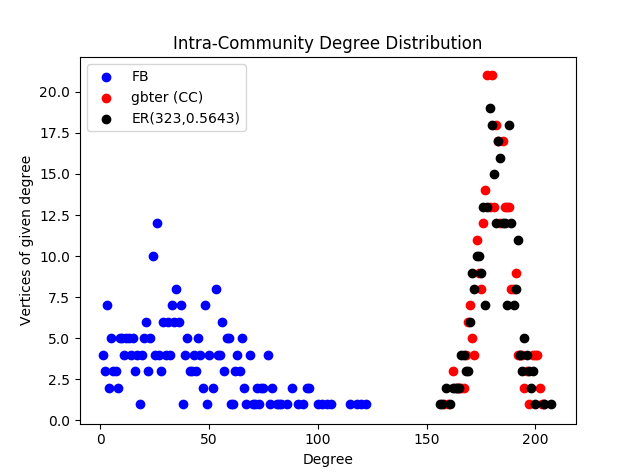}
    \caption{\small{Particular within-community degree distribution comparison using a Facebook ego network\cite{leskovec2012learning}, GBTER (modeling within-community CC), and the corresponding ER process from the GBTER within-community simulation.}}
    \label{fig:gbter_intra_deg}
\end{figure}

\section{Extended GBTER (EGBTER) Model}
\label{sec:4}
\begin{figure}
 \vspace{-.3cm}
    \centering
    \begin{subfigure}[t]{\linewidth}
        \includegraphics[width=\linewidth]{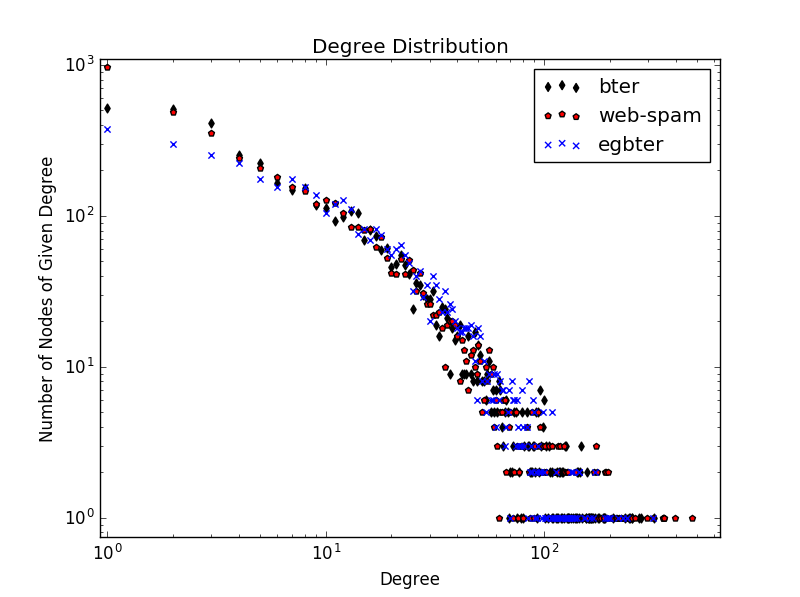}
        \subcaption{\small{Comparison of web-spam \cite{castillo2008web}, BTER and EGBTER degree distributions.}}
        \label{fig:5a}
    \end{subfigure}\hfill
    \begin{subfigure}[t]{\linewidth}
        \includegraphics[width=\linewidth]{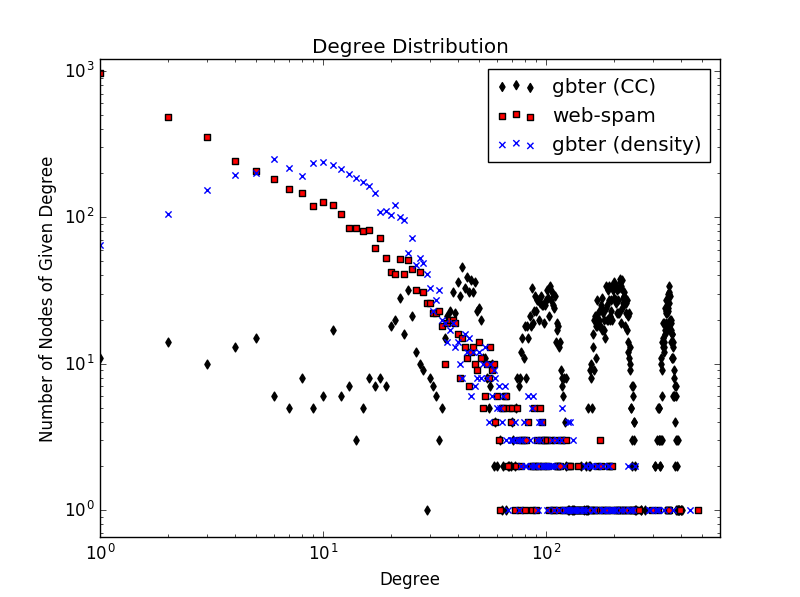}
        \subcaption{\small{Comparison of web-spam and GBTER models degree distributions.}}
        \label{fig:5b}
    \end{subfigure}
     \caption{\small{Note the similar fit of the BTER and EGBTER model in matching the degree distribution for this particular network. GBTER (CC) model's degree distribution is dominated by within-community ER models causing unrealistic fit.}}
     \label{fig:deg_dist_comp1}
\vspace{-.3cm}
\end{figure}

With the goal of capturing degree distribution, CCPD, and membership and density of communities in a single graph model, we propose the EGBTER model. 
Unlike the previous two models, which use ER generation within groupings of nodes (to fulfill CCPD or community density expectations) and CL generation across all nodes (to fulfill the expected degree distribution), the EGBTER model exchanges the within-community ER process (used in the GBTER model) for a BTER analogue. The motivation for doing so is to achieve greater overall accuracy by addressing the deficiencies that may occur at the within-community stage. While GBTER allows modeling of dense sub-regions (communities), the ER processes poorly model within-community degree distributions, as well as clustering coefficients; for example, see Figs.~\ref{fig:gbter_intra_deg},  \ref{fig:deg_dist_comp1}, \ref{fig:6}. The critical insight is that (a) degree distribution and CCPD is important in modeling graphs and these metrics are heavily influenced by edge generation within tight-knit communities, (b) neither are modeled well by ER (and therefore not by GBTER), but (c) both are modeled well by BTER at the graph level; hence, we hypothesize that using a BTER process inside each specified community (and CL process between communities) will give better overall modeling of degrees, CCPD, and community membership and density. Explicit details are given below. \\
\noindent \textit{Inputs:}
\begin{enumerate}[leftmargin=.6cm]
    \item A vertex partition $P=\dot\bigcup_k C_k$ (community membership specified)
    \item Expected within-community degree of each node, $d_i$
    \item Expected global degree of each node, $D_i$, with $d_i\leq D_i$
    \item Within-community CCPD distribution $\lbrace cc_{d,C_k} \rbrace$ for all $C_k$
\end{enumerate}
\textit{Generative process:}
\begin{enumerate}[leftmargin=.6cm]
    \item For each community, $C_k$, place nodes of global degree $D$ into within-community groupings of size $D+1$.
    \item For each within-community grouping, store $p=\sqrt[3]{cc_d}$ for ER process. Here, $cc_d$ is the average CCPD of nodes of degree $d$, where $d$ is the minimum within-community degree of the grouping.
    \item Compute and store the  expected within-community and between-community excess degrees,  
    $\epsilon_i= \max \lbrace{0, d_i-p_L(D)} \rbrace$ and $E_i=D_i-d_i$, respectively. 
    Community density, $p_L$, is the ER probability, and $D$ is the number of neighbors of $v_i$ in this grouping.
    \item Within each community $C_k$ run a BTER process\textemdash ER with parameter $p_k$ and CL with node weight $\epsilon_i$ for all nodes $i\in C_k$. 
    \item For all nodes $v_i\in V$,  run global CL process with node weight $E_i$.  
\end{enumerate}


\section{Experiment}
We compare the BTER, GBTER and EGBTER models on faithful preservation of degree distribution, CCPD distribution community structure for seven real-world networks.  
For degree distribution and CCPD evaluation, the root mean squared error is computed  against the original graph's corresponding feature sequence. 
For community structure, we compare modularity and present a few telling visualizations of real and simulated graphs.  
Modularity is a graph metric taking values in  $[-0.5, 1] $ with higher scores indicating stronger community structure in relation to the given partition; i.e, high modularity indicates high density of edges within communities and sparse edges across communities. 
Borrowing vocabulary from the clustering literature \cite{halkidi2001clustering} modularity is internal validation, i.e., unsupervised, metric.  
Modularity is computed after designating the community structure of the simulated network using the algorithm of Louvain et al.~\cite{blondel2008fast}.  
See Table \ref{tab:notation} for metric formulas. For each metric, we macro-averaged results across 100 simulations for each model using the same input parameters.

  

{\bf Real-world networks description:}
All networks used in this work for model evaluation can be found online either from the Network Repository \cite{nr} or from the Stanford Network Analysis Project (SNAP) \cite{snapnets}. 
Two networks come from biology (a protein interaction network \cite{singh2008-isorank-multi} and a fly brain network), 
two are social networks (a Facebook friend network \cite{leskovec2012learning} and physics collaboration network  \cite{leskovec2007graph}), 
two are web networks \cite{castillo2008web, Boldi-2011-layered, boldi2004-ubicrawler} and one is an infrastructure network \cite{watts1998collective}. 
Only the two social networks come from SNAP.

{\bf BTER model generation:} For the BTER model, we use the online Matlab implementation \cite{feastpack} provided by BTER authors, Kolda et al.~\cite{seshadhri2012community,kolda2014scalable,seshadhri2013triadic}. Step-by-step instructions for fitting BTER to real-world networks are provided online \cite{feastpack}. The supplied code measures/estimates degree and CCPD distributions before sampling edges for the ER and CL processes. We use the default value of any additional parameters. The edge list of each network was exported to Python where graph metrics were computed using the Networkx package \cite{hagberg-2008-exploring}.

{\bf GBTER model generation:} By design, GBTER admits ample flexibility of configuration, in particular, in defining communities members and prescribing within-community ER probabilities. 
Bridges et al. configured the model by learning $p_k$ as within-community density of community $C_k$, but also mentioned that within-community CC could be modeled by setting $p_k = cc_K^{1/3}$.  
We configure GBTER with each of the above two possibilities. 
To partition vertices into communities 
(for GBTER and EGBTER fitting), community assignments are determined by using the Louvain community detection algorithm~\cite{blondel2008fast} in Python~\cite{louvain2016package}.

For each real-world network, a Networkx graph object $G$ is created. 
We find $\lbrace C_k\rbrace$ with Louvain's algorithm, and for each $C_k$ we measure/compute density and average local CC on $G[C_k]$. 
We also computed $\epsilon_i = \max \lbrace  0, d_i - p_k ( \vert C_k\vert -1 ) \rbrace$ for each node $v_i$. 
For edge insertion, we iterate over all possible edges and add the edge with probability given by Eqn.~\ref{eqn:gbter}.

{\bf EGBTER model generation:} 
To generate an EGBTER network,  
we measure inputs from the original network, and for each $C_k$ we generate and store the BTER community groupings, $A_L$,  using global degrees, $D_i$, to determine $A_L$. 
In each $A_L$ we use $p_L = \sqrt[3]{cc_d}$ of the node with lowest within-community degree ($\{d_i\}$)  in $A_L$  to compute $\lbrace \epsilon_i\rbrace$ (Refer to Sec~\ref{sec:4}). 
We then sample edges as done in the scalable BTER model implementation \cite{kolda2014scalable}. 

\begin{table}
\vspace{-.1cm}
\centering
  \caption{Color hierarchy (best to worst) is \colorbox{RoyalBlue!90}{blue}, \colorbox{RoyalBlue!35}{light blue}, \colorbox{red!45}{light red}, \colorbox{BrickRed}{red}, for each network \& metric. 
   Across all networks, BTER is usually first or second  for degree \& CC distributions, yet usually poor in modularity; EGBTER is usually best or second in all three metrics. All metrics are macro-averaged across 100 random realizations for each model.} 
  \label{tab:results}
  \begin{threeparttable}
    \begin{tabular}{ccccc}
    \toprule
       & Model &  $RMSE \lbrace d \rbrace$ & $Q$ &  $RMSE\lbrace cc_{d} \rbrace$ \\
    \midrule
    \multirow{5}[1]{*}{\begin{sideways}bio-dmela\end{sideways}} & true  & NA & 0.4530 & NA \\
          & bter 
          &\cellcolor{RoyalBlue!90}  59.3526 & \cellcolor{red!45} 0.3445 & \cellcolor{RoyalBlue!35}0.0113 \\
          & gbter
          & \cellcolor{red!45} 112.1429 & \cellcolor{RoyalBlue!35} 0.4335 & \cellcolor{RoyalBlue!90} 0.0105 \\
          & gbter$^{\text{CC}}$
          & \cellcolor{BrickRed} 127.8613 &\cellcolor{BrickRed} 0.7087 &\cellcolor{red!45} 0.0118 \\
          & egbter
          &  \cellcolor{RoyalBlue!35}61.7373 & \cellcolor{RoyalBlue!90} 0.4620 & \cellcolor{BrickRed} 0.0181 \\
    \midrule
    \multirow{5}[2]{*}{\begin{sideways}facebook\end{sideways}} & true  &  NA & 0.8350 & NA \\
          & bter 
          & \cellcolor{RoyalBlue!90}2.4670 & \cellcolor{red!45} 0.7160 & \cellcolor{RoyalBlue!90}0.0973 \\
          & gbter
          &\cellcolor{red!45} 10.2906  &\cellcolor{BrickRed} 0.6648  & \cellcolor{red!45}0.2021 \\
          & gbter$^{\text{CC}}$
          & \cellcolor{BrickRed} 16.7792 & \cellcolor{RoyalBlue!35}0.8728 
          & \cellcolor{BrickRed} 0.2617 \\
          & egbter
          & \cellcolor{RoyalBlue!35}3.9274 & \cellcolor{RoyalBlue!90}0.8448 & \cellcolor{RoyalBlue!35}0.1748 \\
    \midrule
    \multirow{5}[2]{*}{\begin{sideways}bn-fly\end{sideways}} & true  & NA & 0.4188 & NA \\
          & bter  
          &\cellcolor{RoyalBlue!90}5.1272 & \cellcolor{red!45}0.5565 
          & \cellcolor{RoyalBlue!35}0.0358 \\
          & gbter 
          & \cellcolor{red!45} 15.7820 & \cellcolor{RoyalBlue!90}0.4497 &\cellcolor{red!45} 0.0464 \\
          & gbter$^{\text{CC}}$
          & \cellcolor{BrickRed} 18.4049 &\cellcolor{BrickRed} 0.6520 
          & \cellcolor{BrickRed} 0.0799 \\
          & egbter 
         & \cellcolor{RoyalBlue!35}7.4551 & \cellcolor{RoyalBlue!35}0.5149 &  \cellcolor{RoyalBlue!90}0.0307 \\
    \midrule
    \multirow{5}[2]{*}{\begin{sideways}ca-GrQc\end{sideways}} & true & NA & 0.8628 & NA \\
          & bter  
          & \cellcolor{RoyalBlue!90}75.9245 & \cellcolor{RoyalBlue!35}0.8093 & \cellcolor{RoyalBlue!90}0.3202 \\
          & gbter 
          & \cellcolor{red!45}84.0651 & \cellcolor{BrickRed} 0.6567 & \cellcolor{red!45}0.4004 \\
          & 
          gbter$^{\text{CC}}$ & \cellcolor{BrickRed} 111.2112 &\cellcolor{red!45} 0.9237 & \cellcolor{BrickRed} 0.4688 \\
          & 
          egbter 
          & \cellcolor{RoyalBlue!35}82.4652 & \cellcolor{RoyalBlue!90}0.8617 & \cellcolor{RoyalBlue!35}0.3476 \\
    \midrule
    \multirow{5}[2]{*}{\begin{sideways}inf-power\end{sideways}} & true  & NA & 0.9357 & NA \\
          & bter 
         & \cellcolor{RoyalBlue!90}151.1312 &\cellcolor{BrickRed} 0.7322 & \cellcolor{RoyalBlue!35}0.0461 \\
          & gbter
          & \cellcolor{RoyalBlue!35}184.9929 & \cellcolor{red!45} 0.7711 
          & \cellcolor{BrickRed}0.0755 \\
          & gbter$^{\text{CC}}$ 
          & \cellcolor{BrickRed}236.2863 & \cellcolor{RoyalBlue!35}0.9266 
          & \cellcolor{red!45} 0.1990 \\
          & egbter 
          & \cellcolor{red!45} 190.5787 & \cellcolor{RoyalBlue!90}0.9399 & \cellcolor{RoyalBlue!90}0.0346 \\
    \midrule
    \multirow{5}[2]{*}{\begin{sideways}web-spam\end{sideways}} & true & NA & 0.5002 & NA \\
          & 
          bter & \cellcolor{RoyalBlue!90}21.3190 & \cellcolor{RoyalBlue!90} 0.5310 & \cellcolor{RoyalBlue!35}0.0590 \\
          & 
          gbter & \cellcolor{red!45} 48.7568 & \cellcolor{RoyalBlue!35} 0.4529 &\cellcolor{red!45} 0.0984 \\
          &
          gbter$^{\text{CC}}$ & \cellcolor{BrickRed} 57.9175 & \cellcolor{BrickRed} 0.7656 & \cellcolor{BrickRed} 0.2235 \\
          &
          egbter
          &\cellcolor{RoyalBlue!35} 29.1909 & \cellcolor{red!45} 0.5693 
          & \cellcolor{RoyalBlue!90}0.0566 \\
    \midrule
    \multirow{5}[2]{*}{\begin{sideways}webbase-2001\end{sideways}} & true & NA & 0.9354 & NA \\ 
          & bter 
          & \cellcolor{RoyalBlue!90}126.9450 &\cellcolor{red!45} 0.7437 & \cellcolor{RoyalBlue!35}0.0494 \\
          & gbter 
          & \cellcolor{red!45}178.8203 &\cellcolor{RoyalBlue!35} 0.7868  & \cellcolor{red!45} 0.0600 \\
          & gbter$^{\text{CC}}$
          &\cellcolor{BrickRed} 225.2118 &\cellcolor{BrickRed} 0.4467  & \cellcolor{BrickRed}0.2136 \\
          & egbter
          & \cellcolor{RoyalBlue!35}134.7270 & \cellcolor{RoyalBlue!90}0.9367 
          & \cellcolor{RoyalBlue!90}0.0340 \\
    \bottomrule
    \end{tabular}%
    \end{threeparttable}
    \vspace{-.2cm}
\end{table}%

For each edge, we sample from a trinomial distribution to determine which process generated the edge (ER, within-$C_k$ CL, or global CL process). 
Weights for each CL process are given by the sum of the excess expected degree sequence divided by total weight for all three processes. 
In the ER process, sampling gives rise to duplicates so in each grouping $A_L$, if we wish to add $\binom{\vert A_L\vert}{2}p_L$ distinct edges, we must perform
$ w(A_L) = \binom{\vert A_L\vert}{2} \log \{(1-p_L)^{-1}\}$  
samples in expectation. See the work of Kolda et al. \cite{kolda2014scalable} for a proof. The ER process total weight is $\sum_{A_L} w(A_L)$ divided by total weight across all three processes. 

Once we determine which process generated the edge, we proceed to sample edge endpoints according to the corresponding edge criteria for that process. We repeat the weighted calculation above but on the level of $A_L$, $C_k$ or globally depending on predetermined edge process. For example, suppose in the first step we determine that our edge is generated in some $ER$ process. Next we sample $A_L$ (a group of nodes) with $P(A_L) \propto w(A_L)$. 
Once $A_L$ is determined, we sample two endpoints, without replacement, uniformly from $\{v_i\in{A_L}\}$ since the ER process has uniform edge insertion probability.




\section{Results}
\label{sec:5}
Table~\ref*{tab:results} summarizes the numerical output of our simulations.



{\bf Degree Distribution Results:}
We see that the BTER model performs the best in $RMSE \lbrace d\rbrace$ across all seven networks and the EGBTER model performs second best in six of seven networks. 
Configuring GBTER to model within-community CC in each community produces the worst error across all networks. 
This is because for the networks used, within-community CC exceeded within-community density, causing the model to overstate the number of edges in each community in the ER process. 
When this happens, the local surplus in each community translates to a global degree distribution that is a superposition of the overstated ER degree distributions from each of the communities. This arises from the implicit assumption in GBTER that communities are well-modeled internally by an ER model. See Fig.~\ref{fig:gbter_intra_deg} and Fig.~\ref{fig:5b} for visualizations of GBTER degree distributions when within-community CC exceeds within-community density.



\begin{figure}
    \vspace{-.3cm}
    \centering 
    \begin{subfigure}[t]{\linewidth}
        \includegraphics[width=\linewidth]{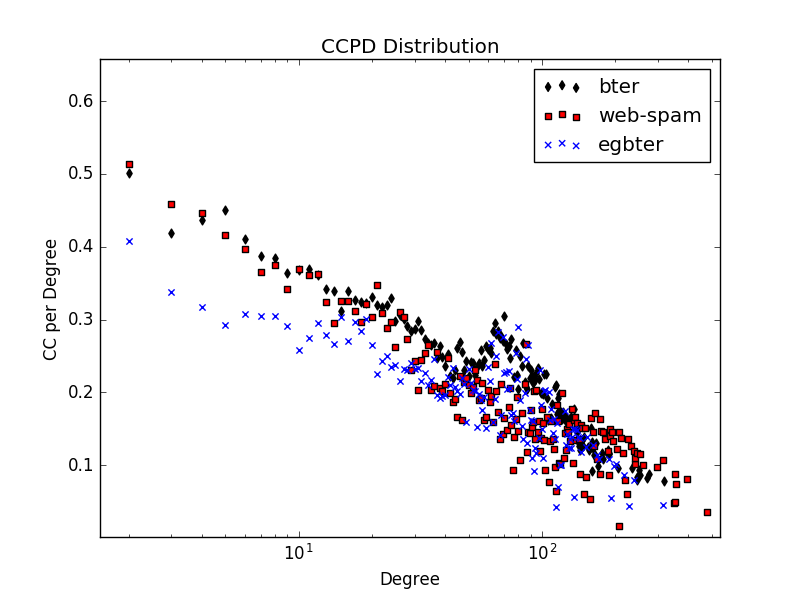}
        \subcaption{\small{Comparison of web-spam \cite{castillo2008web}, BTER and EGBTER CCPD distributions.}}
        \label{fig:6a}
    \end{subfigure}\hfill
    \begin{subfigure}[t]{\linewidth}
        \includegraphics[width=\linewidth]{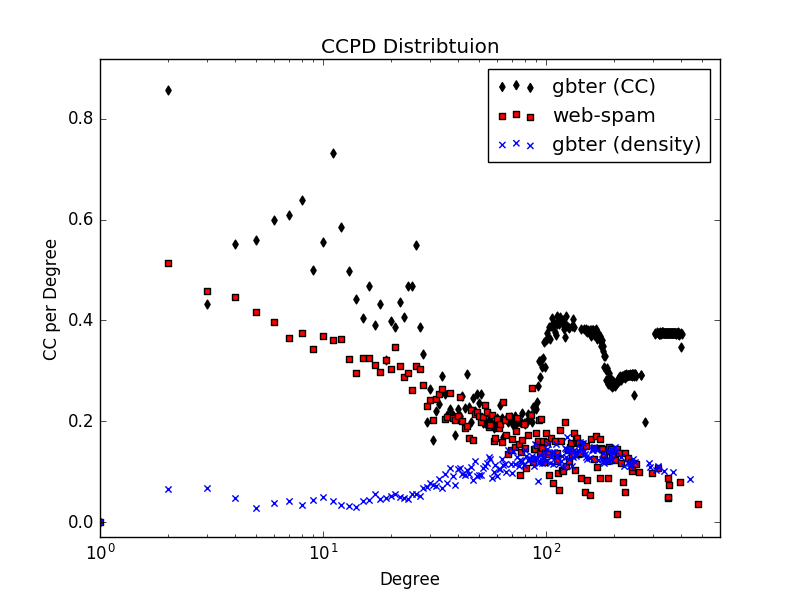}
        \subcaption{\small{Comparison of web-spam and GBTER models' CCPD distributions.}}
        \label{fig:6b}
    \end{subfigure}
       \caption{\small{Visual depiction of web-spam~\cite{castillo2008web} degree distribution vs. simulated analogs. }}
      
    \label{fig:6}
    \vspace{-.3cm}
\end{figure}

{\bf Clustering Coefficient per Degree Distribution Results:}
EGBTER does the best in four of seven networks on minimizing $RMSE\lbrace cc_d \rbrace$ (global CCPD distribution). This is to be expected as we incorporate an analog of the BTER process on the community level to adhere to the within-community CCPD distribution. Despite incorporating community structure into our model, we gain a slight advantage in $\lbrace cc_d \rbrace$. This is not quite clear from Fig.~\ref*{fig:6a}, however is evident from the numerical computation (Table~\ref{tab:results}).

The deficiency of GBTER in modeling $\lbrace cc_d \rbrace$ can be seen in Fig.~\ref*{fig:6b}. The CCPD of high degree nodes is grossly overstated. As with the previously observed edge surplus (Fig.~\ref{fig:5b}), we attribute this to the ER process used within communities in the GBTER model.

\begin{figure*}
    \vspace{-.3cm}
    \centering
    \begin{subfigure}[t]{0.32\linewidth}
        \includegraphics[width=\linewidth]{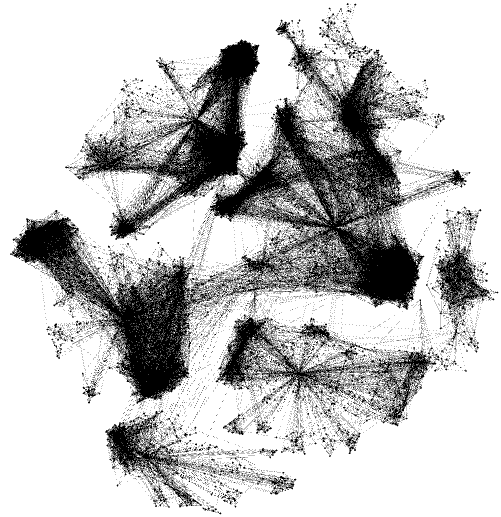}
        \subcaption{\small{Original FB network.}}
        \label{fig:7a}
    \end{subfigure}\hfill    
    \begin{subfigure}[t]{0.32\linewidth}
        \includegraphics[width=\linewidth]{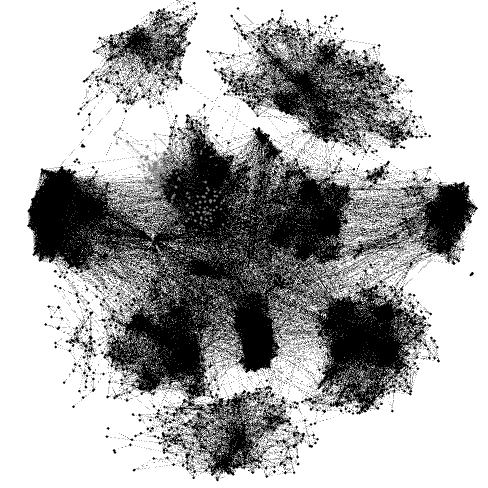}
        \subcaption{\small{EGBTER FB simulation.}}
        \label{fig:7b}
    \end{subfigure}\hfill
    \begin{subfigure}[t]{0.32\linewidth}
        \includegraphics[width=\linewidth]{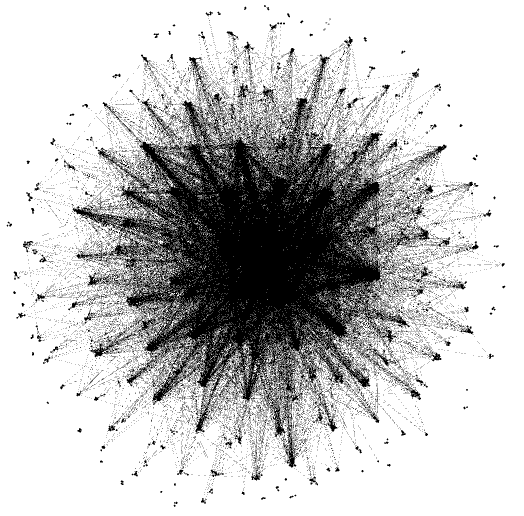}
        \subcaption{\small{BTER FB simulation.}}
        \label{fig:7c}
    \end{subfigure}\hfill \\
    \begin{subfigure}[t]{0.32\linewidth}
        \includegraphics[width=\linewidth]{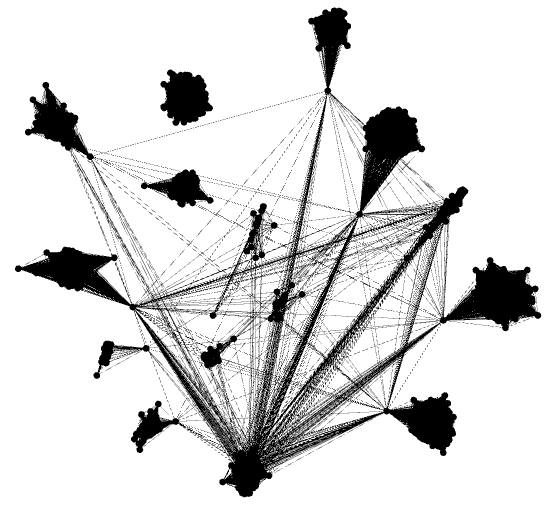}
        \subcaption{\small{GBTER FB simulation (fitting model to within-community CC).}}
        \label{fig:7d}
    \end{subfigure}\hfill
    \begin{subfigure}[t]{0.32\linewidth}
        \includegraphics[width=\linewidth]{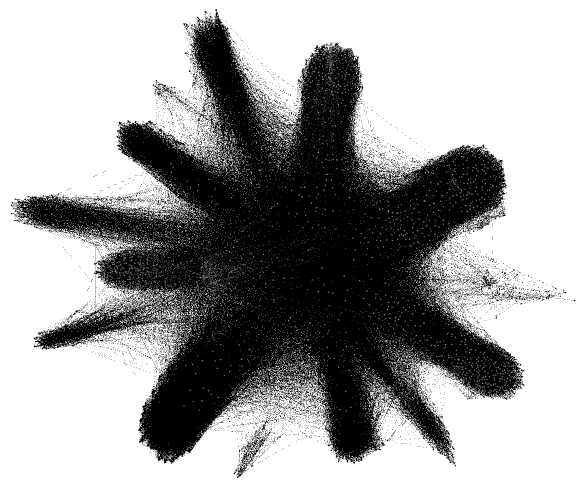}
        \subcaption{\small{GBTER FB simulation (fitting model to within-community density).}}
        \label{fig:7e}
    \end{subfigure}\hfill
    \begin{subfigure}[t]{0.32\linewidth}
        \includegraphics[width=\linewidth]{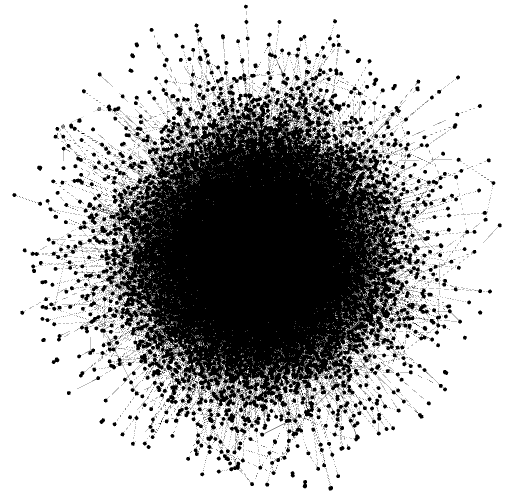}
        \subcaption{\small{bio-dmela BTER visualization.}}
        \label{fig:7f}
    \end{subfigure}
    \caption{\small{Visualizations of FB network and simulated networks according to model. Following quantitative modularity results, EGBTER qualitatively simulates community structure best. We also include a BTER visualization of a biological protein interaction network (f), as BTER tended to exhibit either a star pattern (c) or a dense core (f). All visualizations were produced in Gephi~\cite{ICWSM09154}.}}
    \label{fig:7}
    \vspace{-.3cm}
\end{figure*}

{\bf Community Structure Results:}\\
On modularity, EGBTER performed the best in five of seven networks with GBTER (modeling within-community density) coming in second. This indicates that EGBTER does better in preserving the overall level of community structure in the network than the other models. 

Note that when modeling within-community CC, GBTER tends to overstate the overall level of network division while modeling within-community density tends to have the opposite effect. This agrees with previous discussion on edge metrics where we saw that the degree distribution was a superposition of ER degree distributions from each community when modeling within-community CC. See Fig.~\ref*{fig:7} for network visualizations.

The BTER model does not perform as well as EGBTER and GBTER in preserving modularity, only performing better than the GBTER model when modeling within-community CC. 
This is unsurprising as it does not allow user prescription of community structure. 
Visualizations (using the software Gephi~\cite{ICWSM09154}) of the outputs generated by the BTER model tend to exhibit an organized star-shaped collection of dense ER graphs together with, in some cases, a large dense core (see Fig.~\ref{fig:7c} and Fig.~\ref{fig:7f}). 
The visualization BTER produced depended heavily on which subprocess generated the majority of edges in the underlying simulated network. 
The star-shaped collection arose when the ER process was the dominant edge generation process, while the dense core arose when the CL model was the dominant edge generation process. 
Each distinct type of visualization agrees with previous work on both the BTER model~\cite{seshadhri2012community} and on average distances in CL graphs~\cite{chung2002average,chung2004average}. 
\section{Conclusion}
\label{sec:6}
The EGBTER model presented here showed strong overall performance across metrics measuring degree distribution, clustering coefficients, and community structure.  The BTER model uses an automated grouping process that does well to match both degree and CCPD distribution but fails to preserve community structure at the coarsest scale. This problem is addressed in the GBTER model which imposes community membership and size. However, this comes at a cost to CCPD, as was found to be the case in both configurations of GBTER tested here.

This work was concerned with simultaneous faithful preservation of particular graph characteristics. Performance and scalability are out of the scope of our consideration. Testing showed that in networks with high modularity, internal and global degree and clustering coefficients were nearly identical, except for high degree nodes.
Recent work \cite{seshadhri2013triadic,kolda2014counting} has shown CCPD distribution can be estimated via sampling and scales to large networks. Future research is needed to uncover explicit relationships and network properties that would allow for autonomous and scalable generation of EGBTER graphs without the direct reliance on the measured inputs from a seed graph.
Overall, we hope our contributions will enhance the modeling and analysis capabilities graph theory brings to the many diverse applications. 

\section*{Acknowledgements}
This research was supported in part by the Integrated Joint Cybersecurity Coordination Center (iJC3) Cyber R\&D program, of the U.S. Department of Energy.  Any opinions, findings, conclusions, or recommendations expressed in this material are those of the authors and do not necessarily reflect those of the sponsor of this work.

\section*{Appendix}
\addcontentsline{toc}{section}{Appendix}
\label{app:appendix}

Here we describe the metrics used to describe how well different graph models preserve important network characteristics. The three metrics we use are root mean squared error (RMSE) of both degree distribution and CCPD distribution, as well as modularity as a measure of community structure.

{\bf Degree Distribution:} To measure how well the degree distribution of a simulated network fits the degree distribution of the original network, we use the root mean squared error (RMSE) applied to the simulated degree distribution using the degree distribution of the original network as a baseline.

{\bf CCPD Distribution:}
We apply the RMSE of the CCPD distribution in the same fashion as above. This measure implicitly depends on the set of all degrees produced in the simulated network as we measure the error from the set of ordered pairs $\lbrace (d,cc_d) \rbrace$ in the simulated network to the original network.



{\bf Modularity:}
To evaluate how well a particular vertex clustering (or in our case a partition) captures the community structure of a network, we shall primarily use the modularity metric formulated by Newman \cite{newman2004finding}. Modularity, denoted $Q$, is given by, $Q=\sum_i{e_{ii}-a_i^2},$
 where $e_{ij}$ is the fraction of total edges that connect nodes in community $C_i$ to community $C_j$ and $a_i=\sum_j{e_{ij}}$. We use this metric because it considers the vertex partition as a whole and allows us to compare different partitions of the same network. Modularity gives a sense of the overall level of community structure present in a network relative to the given partition. This is not to say that higher modularity means better community structure but that higher modularity indicates more overall division into communities with respect to the partition used.

{\bf ER Model:}
The Erd{\H{o}}s-R{\'e}nyi (ER) model \cite{erdos1959random} was one of the first random graph models. The ER model takes as input, the desired number of nodes $n$ and an edge insertion probability $p$. Then a graph on $n$ nodes is generated by inserting each of the $\binom{n}{2}$ possible edges with probability $p$. This graph is denoted as $ER(n,p)$. As a result of this simple insertion process, in expectation we have network density, $cc_{v_i}$ and CCPD equal to $p$. We also have $\binom{n}{2}p$ edges in expectation and \begin{equation} P(d_i=k)=\binom{n-1}{k}p^{k}(1-p)^{n-1-k}. \end{equation} This produces a roughly Gaussian shaped degree distribution, counter to that observed in real world networks \cite{barabasi1999emergence}.\\

{\bf CL Model:}
The Chung-Lu (CL) model \cite{aiello2001random} arose as an attempt to model graphs with a power law degree distribution. In the CL model, each node $v_i$ is assigned a weight, $w_i$. The CL model takes as input, the list of corresponding vertex weights $\lbrace w_i \rbrace$. The probability of inserting edge $v_iv_j$ is
\begin{equation}
    P(v_iv_j\in{E})=\frac{w_iw_j}{\sum_{v_k\in{V}}w_k},
    \label{eq:6}
\end{equation} which we shall denote by $CL(w_i,w_j)$.
For this probability to be well defined we require that $w_iw_j\leq \sum_{v_k\in{V}}w_k$ for all $v_i,v_j\in{V}$. It is possible that the RHS of Eqn. \ref*{eq:6} is greater than 1. For our purposes, if this occurs we set $P(v_iv_j\in{E})=1$. A common adaptation is to use $w_i=d_i$ for all $v_i\in{V}$. When $w_i=d_i$, this is referred to as a null model. CL generalizes ER and retains independent edge insertion while being able to match $d_i$ in expectation. However, this process rarely closes triangles \cite{kolda2014scalable} and the average distance is relatively small \cite{chung2002average,chung2004average}. This makes this model a bad fit for CC and community structure but laid a foundation for capturing network degree distribution in expectation. See Fig.~\ref*{fig:7f} for BTER visualization of a biological network. This visualization looks very similar to those of CL models.



\small
\bibliographystyle{IEEEtran}
\bibliography{IEEEabrv,refs}


\end{document}